\documentclass[fleqn,twoside]{article}
\usepackage{espcrc2}
\usepackage{graphicx}
\usepackage{epsf}

\newcommand{\reaction}{$\Lambda_b^0\to\Lambda_c^+l^-\bar{\nu}_l$}
\newcommand{\lb}{$\Lambda_b$}

\newcommand{\qc}{$c$}
\newcommand{\qb}{$b$}

\begin{document}

\title{Study of the semileptonic decay \reaction}

\author{
C. Albertus
\address[Granada]{Departamento de F\'\i sica Moderna, 
Facultad de Ciencias, Universidad de Granada.\\ E-18071, Granada, Spain.},
E. Hern\'andez
\address{Grupo de F\'\i sica Nuclear, 
Facultad de Ciencias, Universidad de Salamanca.\\ E-37008, Salamanca, Spain.},
J. Nieves \addressmark[Granada]
\thanks{This work is partially supported by DGI and FEDER funds, under contracts 
BFM2002-03218 and BFM2003-00856, and by Junta de Andaluc\'\i a and Junta de Castilla
y Le\'on under contracts FQM0225 and SA104/04.  C. A. wishes to acknowledge 
a grant related to his PhD from Junta de Andaluc\'\i a.}}

\begin{abstract}
  Within the  framework of a  nonrelativistic quark model  we evaluate
  the six form factors associated to the \reaction\ semileptonic decay.
  The  baryon  wave  functions  were evaluated  using a  variational
  approach applied to a family of trial functions constrained by Heavy Quark
  Symmetry (HQS).  We use a spectator model with only one-body current operators.
  For these operators we keep up to first order terms on the internal (small) heavy
  quark momentum, but  all  orders on the transferred (large) momentum.
  Our result for the partially integrated decay width is in good agreement 
  with lattice calculations.  Comparison of our total decay width to experiment
  allows us to extract the 
  $V_{cb}$ Cabbibo-Kobayashi-Maskawa matrix element for which we obtain a value
  of 
  $|V_{cb}|=0.047\pm 0.005$ in agreement with a recent determination by the
  DELPHI Collaboration. 
  Furthermore, we  obtain 
  the  universal Isgur-Wise  function with a slope parameter $\rho^2=0.98$ in
   agreement with lattice results. 
\end{abstract}
\maketitle
\section{INTRODUCTION}

Since the discovery of the \lb\ baryon at CERN \cite{al91}, and the discovery
of
most of the charmed baryons \cite{pdg02} of the SU(3) multiplet on the
second level of the SU(4) 20-plet, a great deal of theoretical work has been
devoted to their study
\cite{bow98}-\cite{dun98}.


On the other hand, HQS has shown itself as an
excellent tool to understand charm and bottom physics ~\cite{neu94}. It has
extensively 
been used to describe systems containing a heavy quark (\qc\
or \qb), being, for instance,  one of the basis in lattice QCD
simulations of bottom systems.
HQS is an approximate SU($N_F$) symmetry of QCD, being $N_F$ the number of
heavy flavours, which appears in  systems containing heavy quarks with masses
much larger than any other energy scale ($q$ = $\Lambda_{QCD}$, $m_u$,
$m_d$, $m_s$,\ldots) controlling
the dynamics of the remaining degrees of freedom. For baryons containing a heavy
quark, and up to
corrections of the order\footnote{Here $q$ stands for a typical energy scale
relevant for the light degrees of freedom while $m_h$ is the mass of the heavy
quark} ${\cal O}(\frac{q}{m_h})$,  HQS
guarantees that the heavy baryon light degrees of freedom quantum
number are always well defined.

However, HQS has not been systematically used within the context of 
nonrelativistic constituent quark models (NRCQM). Very recently,
we have proposed~\cite{aahn03} a simple method to solve the nonrelativistic three body 
problem for baryons with a heavy quark, where we have made full use of the
consequences of HQS for that systems.
Thanks to HQS, the method proposed provides us with 
simple wave
functions, while the results obtained for the spectrum and other observables
compare quite well with
more sophisticated Faddeev calculations done in \cite{si96}.

The purpose of the present work is the calculation of the
semileptonic decay \reaction\ within the context of NRCQM and
HQS by making use of the wave functions obtained in Ref.~\cite{aahn03}.

This manuscript is organized as follows: In section 2 we  provide a general
overview of the calculational details needed to evaluate the different
observables. In section 3 we give some preliminary results and, finally, the
 conclusions are presented in section 4.

\section{THE MODEL}

\subsection{BARYON WAVE FUNCTION}
\begin{figure}
\vspace{-4cm}
\centerline{\includegraphics[height=25cm]{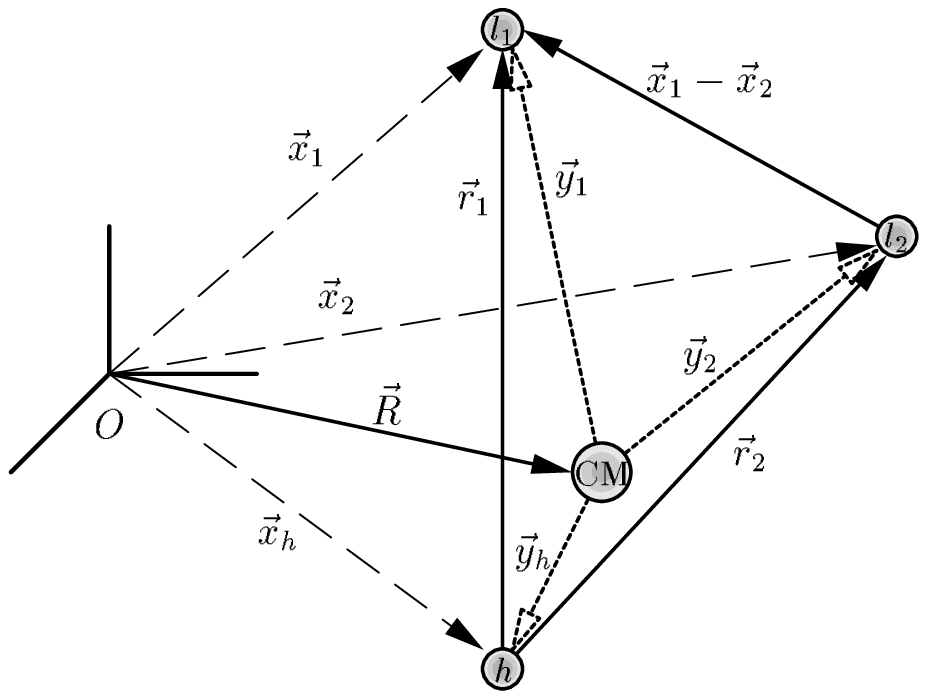}}
\vspace{-15cm}
\vspace{-1cm}\caption{ Definition of different coordinates used.
}
\label{fig:coor}
\end{figure}
Working with the coordinates $\vec{R}$, $\vec{r}_1$ and $\vec{r}_2$
(see Fig.~\ref{fig:coor}) we can separate the centre of mass motion
and the Hamiltonian of the three quark system reads
\begin{equation}
H=-\frac{\vec\nabla_{\vec{R}}^2}{2 M} + H^{\rm int} 
\end{equation}
\begin{eqnarray}
H^{\rm int}&=&\sum_{i=1,2}\ H_i^{sp}+V_{l_1l_2}(\vec{r}_1-\vec{r}_2,spin)\nonumber\\&&-
\frac{\vec\nabla_1\cdot\vec\nabla_2}{m_h}+M\nonumber\\
H_i^{sp}&=&-\frac{\vec\nabla_i^2}{2\mu_i}+V_{hl_i}(\vec{r}_i,spin), \ \ i=1,2
\end{eqnarray}
The suffix $h$ stands for the heavy quark while $l_{1,2}$ refer to the light
quarks. $M=(m_{l_1}+m_{l_2}+m_h)$, $\mu_{i}=(1/m_{l_i}+1/m_{h})^{-1}$ and $\vec\nabla_i$ is the gradient with
respect to $\vec r_i$. The internal
wave function for a $\Lambda_h$ baryon ($h=b,c$) with spin projection $s$, 
isospin $I=0$ and total spin 
\hbox{$S_{light}=0$} for the
light degrees of freedom is given by 
\begin{eqnarray}
\left|\Lambda_h; s\right>&=&\left\{\left|00\right>_{Iso.}
\otimes\left|00\right>_{S_{light}}\right\}
\nonumber\\&&\otimes\left|h; s\right>\psi(\vec{r}_1,\vec{r}_2)
\end{eqnarray}
The spatial wave function is obtained by using a simple variational ansatz
\begin{eqnarray}
\psi(\vec{r}_1,\vec{r}_2)&=&\varphi(r_1,r_2,r_{12})\nonumber\\
&=&N\phi_{l_1}^h(r_1)\phi_{l_2}^h(r_2)F(r_{12})
\end{eqnarray}
where $N$ is a normalization factor and $F(r_{12})$ is a Jastrow-type
correlation function
\begin{equation}
F(r_{12}) = \sum_{j=1}^4 a_j e^{-b_j^2(r_{12}+d_j)^2}
\end{equation}
being $a_1=1$ and $a_{i\ne 1},b_i,d_i$ free parameters.
$\phi_{l_1}^h(r_1)$ and $\phi_{l_2}^h(r_2)$ are essentially fixed by
the s-wave ground state wave functions of the single particle
Hamiltonians $H_{1,2}^{sp}$ for the relative motion of a light quark
with respect to the heavy one. These wave functions are corrected at
large distances where modifications coming from the presence of the
other light quark are expected. This modification introduces two extra
variational parameters.


Further details concerning wave functions, two-quark potentials used and the
fixing of the parameters can be
found in Ref. \cite{aahn03}.

\subsection{DECAY WIDTH}
Neglecting  lepton masses the differential cross section can be written as

%
\begin{eqnarray}
\frac{d\,\Gamma}{d\omega}&=&\frac{G_F^2}{12\pi^3}\,|V_{cb}|^2\,M_{\Lambda_c}^3\,q^2
\sqrt{\omega^2-1}\nonumber\\
&& (-g^{\alpha\beta}+\frac{q^\alpha q^\beta}{q^2})\, H_{\alpha\beta}(q)
\end{eqnarray}
where $\omega$ is the product of four velocities 
\hbox{$\omega=(p/M_{\Lambda_b})\cdot (p'/M_{\Lambda_c})$}, $q=p-p'$ and 
$H_{\alpha\beta}(q)$ is the hadronic tensor defined as
\begin{eqnarray}
&&\hspace{-.7cm}H_{\alpha\beta}(q)\nonumber\\
&&\hspace{-.5cm}=\overline{\sum_{s,s'}}\ \langle\Lambda_c;s',\vec p\,'=-\vec{q}\,|
(j_{cc})_\alpha(0)|\Lambda_b;s,\vec{p}=\vec{0}\rangle\nonumber\\
&&\hspace{0cm}(\langle\Lambda_c;s',\vec{p}\,'=-\vec{q}\,|
(j_{cc})_\beta(0)|\Lambda_b;s,\vec{p}=\vec{0}\rangle)^*
\end{eqnarray}
where  $(j_{cc})^\alpha(0)=\bar c(0)\gamma^\alpha(1-\gamma_5)b(0)$, and where
$\vec p$ ($\vec p\,'$) stands for the three-momentum  of the $\Lambda_b$ ($\Lambda_c$) baryon.
We have taken the $\Lambda_b$ baryon at rest and we have 
 averaged (summed) 
 over the spin $s$ ($s'$) of the $\Lambda_b$ ($\Lambda_c$) baryon. Baryon 
 states are normalized to ``\,E/M\,''.
%

The matrix element of the weak charged current 
between hadronic states is parametrized in the usual way
\begin{eqnarray}
&&\hspace{-.5cm}\left<\Lambda_c; s,\vec{p}\,'|(j_{cc})^\alpha(0)|\Lambda_b;
s,\vec{p}\right>\nonumber\\
&&\hspace{.5cm}=\bar{u}_{\Lambda_c}^{(s')}(\vec{p}\,')\left[\ \gamma^\alpha(F_1-\gamma_5G_1)
\right.\nonumber\\
&&\hspace{2.2cm}\left.+v^\alpha(F_2-\gamma_5G_2)\right.\nonumber\\
&&\hspace{2.2cm}\left.+v'^{\alpha}(F_3-\gamma_5G_3)\ \right]u_{\Lambda_b}^{(s)}(\vec{p}\,)
\end{eqnarray}
where $v^\alpha=p/M_{\Lambda_b}$ ($v'^{\alpha}=p'/M_{\Lambda_c}$) is the four-
velocity of the $\Lambda_b$ ($ \Lambda_c$) baryon.\\
In our nonrelativistic calculation we evaluate
%
\begin{eqnarray}
&&\hspace{-.75cm}\sqrt{\frac{M_{\Lambda_b}}{E_{\Lambda_b}(\vec{p})}}
\sqrt{\frac{M_{\Lambda_c}}{E_{\Lambda_c}(\vec{p}\,')}}
\left<\Lambda_c; s,\vec{p}\,'| 
(j_{cc})^\alpha(0)
|\Lambda_b;
s,\vec{p}\right>\nonumber\\
%
\end{eqnarray}
which in momentum space is given by
%
\begin{eqnarray}
\label{momesp}
&&\hspace{-.6cm}
\int d^3q_1d^3q_2 d^3q_h d^3q_h^\prime\ 
\nonumber\\
&&\hspace{-.5cm}\sqrt{\frac{m_b}{E_b(\vec{q}_h)}}\sqrt{\frac{m_c}{E_c(\vec{q}\,'_h)}}
[\bar{u}_c^{(s')}(\vec{q}\,'_h)\gamma^\mu(1-\gamma_5)u_b^{(s)}(\vec{q}_h)]\nonumber\\
&&\phi^{*(c)}(\vec{p}\,';\vec q_1,\vec q_2,\vec q\,'_h)\ 
\phi^{(b)}(\vec{p};\vec q_1,\vec q_2,\vec q_h)
\end{eqnarray}
with $\vec p\,'=\vec p-\vec q$. The wave functions in momentum space appearing in the above equation 
are the Fourier
transformed of those in coordinate space
%
%
\begin{equation}
\psi_{p}(\vec{x}_1,\vec{x}_2,\vec{x}_h)=
\frac{e^{i\vec{p}\cdot\vec{R}}}{(2\pi)^{\frac{3}{2}}}\psi(\vec{r}_1,\vec{r}_2)
\end{equation}
with $\psi(\vec{r}_1,\vec{r}_2)$  described in the previous
subsection. 

The actual calculations are done in coordinate space. For that we need to expand the
$b\to c$ transition operator in Eq.(\ref{momesp}). In  this expansion we shall keep
up to terms in first order on $\vec q_h$. Being the  $\Lambda_b$ baryon  at rest,
$\vec q_h$ is an internal momenta  which is much smaller than any of the
heavy quark masses. On the other hand the transferred momentum $\vec q$ can be large
so that we do an exact expansion on $\vec q$.\\
\vspace*{.5cm}
\section{PRELIMINARY RESULTS}
\begin{figure}
\begin{center}
\makebox[0pt]{\input{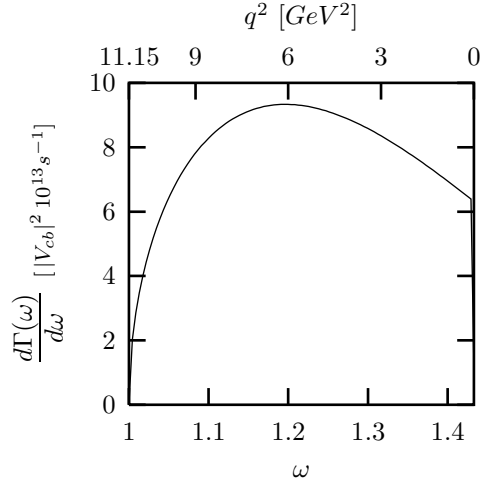}
}\\
\vspace{-.8cm}\caption{Differential decay width}
\label{fig:dif}
\end{center}
\end{figure}
We present here the results obtained with the use of the wave functions that derive
from the AL1 two-quark interaction potential (see Ref.~\cite{aahn03} and references
therein for details).

%
In Fig. \ref{fig:dif} we show the differential decay width $d\Gamma(\omega)/d\omega$
in terms of $\omega$ (lower $x$-axis) and  $q^2$ (upper $x$-axis).

The partially integrated value 
\begin{equation}
\int_1^{1.2}d\omega\ \frac{d\Gamma(\omega)}{d\omega}=1.49\left|V_{cb}\right|^2\cdot10^{13}s^{-1}
\end{equation}
agrees nicely with a previous lattice calculation \cite{bow98} which gives
 for this integral the 
value 
%
$1.4_{-4}^{+5}\left|V_{cb}\right|^2\cdot10^{13}s^{-1}$.
%
Our total width is given by
%
\begin{equation}
\int_1^{\omega_{max}}d\omega \frac{d\Gamma(\omega)}{d\omega}=3.41\left|V_{cb}\right|^2\cdot10^{13}s^{-1}
\end{equation}
Comparing to experimental data in Ref.~\cite{pdg04} we can extract the value
for the CKM matrix element
\begin{equation}
|V_{cb}|=0.047\pm 0.005
\end{equation}
where we quote the error that derives from the experimental uncertainties. This
value is in agreement with the recent determination by the DELPHI
Collaboration~\cite{delpi04}
\hbox{$|V_{cb}|=0.0414\pm 0.0012\pm0.0021\pm 0.0018$}, 
obtained from the analysis of the $\overline{B^0_d}\to D^{*+}l^-\bar{\nu}_l$ reaction.
\begin{figure}
\begin{center}
  \makebox[0pt]{\input{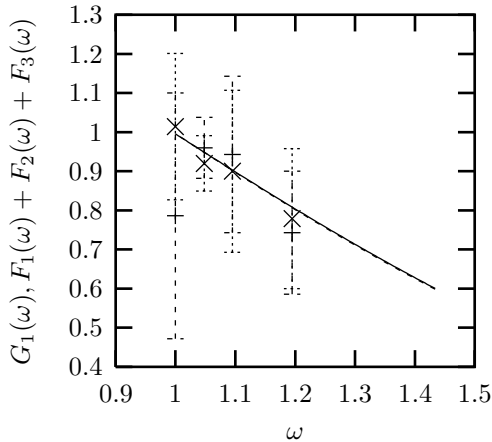}
}\\
\vspace{-.8cm}\caption{Form factor $G_1(\omega)$ (solid line) and the sum $F_1(\omega)
+F_2(\omega)+F_3(\omega)$ (dashed line). Data points are lattice calculations
extrapolated down to the chiral limit as obtained in Ref.~\protect{\cite{bow98}}}
\label{fig:iw}
\end{center}
\end{figure}

In Fig. \ref{fig:iw} we show  the results for the form factor $G_1(\omega)$ and the combination of
form factors $F_1(\omega)+F_2(\omega)+F_3(\omega)$. Both quantities are protected
by Luke's theorem \cite{luke90} from ${\cal O}(1/m_h)$ corrections and  are thus
 very close to the universal Isgur-Wise
 function~\cite{iw91}. As we see from the figure the two quantities are 
 almost identical. We also show in the figure lattice results for the Isgur-Wise
 function obtained 
 in Ref.~\cite{bow98} 
 when the extrapolation to zero light quark
 masses is done.
The value for the slope parameter $\rho^2$ defined as minus the slope at $\omega=1$ is
given by
\begin{equation}
\hspace{1.5cm} \rho^2=0.98
\end{equation}
and it is in good agreement with the central value obtained in the lattice determination 
 which gives
$\rho^2=1.1\pm1.0$.

\section{CONCLUSIONS}
We have presented a calculation of the \reaction \ reaction within the context of NRCQM. 
We use manageable wave functions that were obtained in Ref.~\cite{aahn03} using a simple
variational ansatz based on HQS. 

Our results for the partially integrated decay width and 
the Isgur-Wise function are in good agreement with previous lattice determinations.
Comparison of our total decay width to  experimental data allows us to obtain a 
value for $|V_{cb}|$ in agreement with a recent determination by the DELPHI
Collaboration.


\begin{thebibliography}{99}
\bibitem{al91} C. Albajar {\it et al.,} Phys. Lett. {\bf B273}, 540 (1991); G.
Bari {\it et al.,} Nuovo Cimento {\bf A10}, 1787 (1991)
\bibitem{pdg02} K. Hagiwara {\it et al.,} Phys. Rev.  {\bf D66}, 010001 (2002). 
\bibitem{bow98} K. C. Bowler {\it et al.}, Phys. Rev. {\bf D57} 6948 (1998).
\bibitem{hua01} M-Q Huang, J-P Lee, C. Liu and H.S. Song, Phys. Lett. {\bf B502} 133 (2001).
\bibitem{car00} F. Cardarelli and S. Simula, Nucl. Phys. {\bf A663} 931 (2000);
 Phys. Rev. {\bf D60} 074018 (1999); Phys. Lett. {\bf B421} 295 (1998). 
\bibitem{iva99} M. A. Ivanov, J. G. K\"orner, V. E. Lyubovitskij and A. G.
Rusetsky, Phys. Rev. {\bf D59} 074016 (1999).
\bibitem{lee98} J-P Lee, C. Liu and H.S. Song, Phys. Rev. {\bf D58} 014013 (1998).
\bibitem{dun98} I. Dunietz, Phys. Rev. {\bf D58} 094010 (1998).
\bibitem{neu94} N. Isgur and M. B. Wise, Phys. Lett.{\bf B232}, 113 (1989); 
M. Neubert, Phys. Rep. 245 259 (1994); J. G. K\"orner, M. Kr\"amer and D.
Pirjol, Prog. Part. Nucl. Phys. {\bf 33}, 787 (1994).
\bibitem{aahn03} C. Albertus, J. E. Amaro, E. Hern\'andez, J. Nieves. Nucl. Phys. {\bf A 740} (2004) 333-361
\bibitem{si96} B. Silvestre-Brac, Few-Body Systems {\bf 20} (1996) 1.
\bibitem{pdg04} S. Eidelman {\it et al.}, Phys. Lett. {\bf B592}, 1 (2004).
\bibitem{delpi04} J. Abdallah {\it et al.} (DELPHI Collaboration), Eur. Phys. J. {\bf
C33}, 213 (2004)
\bibitem{luke90} M.E. \ Luke, Phys. Lett. {\bf B252}, 447 (1990).
\bibitem{iw91} N.\ Isgur and M. B.\ Wise, Nucl. Phys. {\bf B348} (1991) 292.
\end{thebibliography}
\end{document}